\documentclass[11pt]{article}

\usepackage[utf8]{inputenc}
\usepackage[T1]{fontenc}
\usepackage[margin=1in]{geometry}
\usepackage{lmodern}
\usepackage{microtype}
\usepackage{amsmath,amssymb,amsthm}
\usepackage{mathtools}
\usepackage{graphicx}
\usepackage{xcolor}
\usepackage{booktabs}
\usepackage{enumitem}
\usepackage[hidelinks]{hyperref}
\usepackage[nameinlink,capitalise]{cleveref}

\theoremstyle{definition}

\theoremstyle{remark}

\title{Adaptive Alarm Threshold Prediction in 4G Mobile Networks: A Percentile-Guided Deep Learning Framework with Interpretable Outputs}
\author{%
  Ayon Roy\thanks{\texttt{ext.ayon.roy@bracu.ac.bd}}
  \and
  Sadman Sharif\thanks{\texttt{sadman.sharif@g.bracu.ac.bd}}
  \and
  Shiva Prasad Sarkar\thanks{\texttt{shiva.prasad.sarkar@g.bracu.ac.bd}}
}
\date{\today}

\begin{document}

\maketitle

\begin{abstract}
In mobile telecommunications, alarms act as early warning signals. They are triggered when a cell, the basic unit of radio coverage, shuts down or behaves abnormally. This signals a degradation in service quality, which directly affects the customer experience. To fix the issue, operators rely on preset thresholds to decide when an engineer should be sent out. In practice, these thresholds are set manually and remain fixed regardless of the time of day, traffic levels, or overall network conditions. This often leads to serious faults slipping through during busy hours, while minor issues can cause unnecessary callouts when the network is quiet. This paper presents a machine learning framework that automatically predicts four alarm thresholds, audit window duration, inactive time limit, total fluctuation count, and per hour fluctuation limit, from live network behavior. Since no ground truth labels exist for thresholds, we introduce a percentile guided label derivation strategy and evaluate four models on an anonymized dataset of 10,648 cells across three vendors and nine regions from a real 4G network, comprising a Gradient Boosted Trees baseline, a CNN-BiLSTM with attention, the proposed PCTN, and an iTransformer. PCTN performs the best overall with respect to three of the four targets, outperforming a state-of-the-art iTransformer while using 83 percent fewer parameters. Its mixed output heads and dynamic alpha mechanism produce thresholds that are both accurate and interpretable, allowing operators to inspect and adjust the learned policy without retraining. All comparisons are statistically significant at p < 0.001. The framework undergoes daily retraining using new data, which enables the thresholds to constantly adjust to changes in the network.
\end{abstract}



\section{Introduction}
Mobile networks are built on thousands of small units called cells. Each cell is in charge of covering a certain area. Customers in that area lose their connection when a cell stops working or keeps turning on and off. Telecom operators rely on alarm systems to keep an eye on how individual cells are behaving and to flag anything that looks off to their engineers. An alarm will trigger based on a threshold, which determines how long a cell can stay inactive, or how many times it can fluctuate, before taking action \cite{costa2009,jakobson1993}. The problem is that these thresholds are set manually and rarely updated. A threshold that works at midnight when the network is quiet is often wrong in the morning when millions of users are active simultaneously. While real faults during busy hours go unnoticed, basic actions during quiet periods can cause engineers to make unnecessary trips \cite{kliger1995}. As networks expand and traffic patterns change, this gap continues to grow \cite{ericsson2023}.

Researchers have investigated the use of machine learning for network fault prediction and alarm classification, which led to promising results \cite{agarwal2022,huang2019}. Sequential models like CNN and LSTM networks have been found to be ideal at learning patterns from time series network data \cite{zhao2021}. Recurrent models have many challenges with long-range dependencies, whereas Transformer models have improved this significantly \cite{vaswani2017}. Liu et al.~\cite{liu2024} recently proposed the iTransformer, which shifts the traditional attention mechanism so that the model learns how features are related to each other instead of time steps. This makes it work well for a wide range of forecasting tasks. Despite these advances in error detection and traffic prediction, none of them address a fundamental question: what should the threshold actually be? Existing methods predict whether a fault will happen or how severe it is. No prior work predicts the decision boundary that determines whether an alarm fires at all. 

This is not a straightforward problem to solve. There are no labels in any network log that say what the correct threshold should have been at a given moment. The label is something that has to be derived from the data itself. To address this, we build a framework that learns four alarm thresholds directly from live network behavior, using a percentile based label strategy that encodes how the network actually acts across different times and days of the week. We also introduce PCTN, which stands for Percentile Guided Contextual Threshold Network, a new model that learns the statistical behavior of the network first and then derives thresholds from those learned patterns rather than predicting numbers directly. This makes the model's reasoning easy to understand and inspect. After evaluating several researches, no prior research has found that conducted adaptive alarm threshold prediction as a machine learning problem or tested it on real data from a live multi-vendor 4G network.

The rest of the research is organized as follows: \Cref{sec:related-work} presents related work, \Cref{sec:dataset-problem} describes the dataset and problem setup, \Cref{sec:proposed-models} presents the models, \Cref{sec:experimental-setup} covers experimental setup, \Cref{sec:results-analysis} reports results, and \Cref{sec:limitations-future,sec:conclusion} discuss limitations, future work, and conclusions.

The primary contributions made by this work are:
\begin{itemize}[leftmargin=1.5em]
\item A formal definition of the adaptive alarm threshold prediction problem and a percentile based label derivation method that creates training supervision from network behavior data where no ground truth labels exist.

\item The PCTN architecture, a new model that predicts the statistical distribution of network behavior first and derives thresholds analytically, producing results that are both accurate and interpretable.

\item An adaptation of the iTransformer to multi output threshold prediction, showing that feature wise attention is well suited to the rich multivariate structure of alarm monitoring data.

\item A four model comparison with statistical significance testing on real anonymized data from a live 4G network covering 10,648 cells across three equipment vendors and nine geographic regions.

\item Identification of the floor-concentration problem in fluctuation targets and a structural solution via the Bernoulli gate, demonstrating ${R^2}$ improvement from -0.34 to 0.57 through explicit binary decomposition of the prediction task.
\end{itemize}

\section{Related Work}
\label{sec:related-work}
 
This section reviews research across three areas that are directly relevant to this work: alarm management in telecommunication networks, fault prediction and anomaly detection using machine learning, and deep learning architectures for time series data.
 
\subsection{Management in Telecom Networks}
 
Alarm management has been an operational challenge in telecom networks for long times. Alarm correlation, a way to group related alarms so that engineers don't have to deal with as many alerts, was first introduced by Jakobson and Weissman~\cite{jakobson1993}. The issue has become more complicated as networks have gotten bigger and more diverse since then. Costa et al.~\cite{costa2009} suggested a smart alarm management system that used rules to sort and prioritize alarms. This demonstrated that automation could significantly minimize the workload of engineers. Recently, Asres et al.~\cite{asres2022} employed machine learning to forecast the likelihood of a network alarm leading to a formal trouble ticket, based on features derived from alarm history within a sliding time window. This is relevant to our work because it shows that alarms can learn how to act in certain ways. However, none of these works address the question of what threshold should have been used to generate the alarm in the first place. They assume the threshold is fixed and try to work around its limitations rather than learning it directly.
 
In today's rapidly changing telecom industry, traditional anomaly detection techniques that depend on rule-based systems are not feasible \cite{edozie2025}. This observation motivates the need for data driven threshold management, which is what this paper addresses.
 
\subsection{Machine Learning for Network Fault Prediction}
 
An emerging area of research has used machine learning to forecast network failures before they lead to service disruptions. Murphy et al.~\cite{murphy2019} performed an extensive analysis on the different approaches to predicting faults in networks and concluded that sequence handling neural networks are the most promising. Huang et al.~\cite{huang2019} demonstrated the effectiveness of machine learning in predictive analytics of operational data on the basis of base station performance indicators by using operational data to forecast future malfunctions. Asres et al.~\cite{asres2022} showed the effectiveness of using alarm history in predictive modeling. These approaches are useful for anticipating problems but they do not help operators decide how sensitive the alarm system should be at a given point in time. Our work is different in that we treat the threshold as the thing to be predicted, not the fault itself. This is a fundamental shift in how the alarm problem is framed, and it is complementary to fault prediction rather than a replacement for it.
 
Anomaly detection in wireless networks has also been studied using unsupervised methods. Several studies have used autoencoders and statistical models to identify unusual network behavior, without needing labeled training data. While these approaches are valuable for exploratory monitoring, they do not produce the kind of actionable threshold values that operators can plug directly into alarm systems.
 
\subsection{Deep Learning Architectures for Time Series Data}
 
Since network alarm data is inherently sequential, the choice of model architecture matters. It has been shown that LSTM and CNN have the ability to detect local feature patterns and temporal relationships in the data in the presence of networks \cite{zhao2021}. The research showed that using convolutional layers to extract features, combined with LSTM layers for standard sequence modeling, produced better results than either approach used on its own. Our CNN BiLSTM model was designed with this understanding in mind.
 
Vaswani et al.~\cite{vaswani2017} proposed the Transformer architecture that introduced attention mechanisms which can model dependencies between any two positions in a sequence regardless of distance. This made Transformers attractive for network time series tasks where relevant patterns may span many hours. However, standard Transformers treat each time step as a token, which means the model attends across time. Liu et al.~\cite{liu2024} contested this design using the iTransformer, which regards each feature as a token and calculates attention across the feature dimension instead. This helps the model learn which features affect each other instead of which time steps are related. This property is particularly useful in our setting, where 123 engineered features from different vendors, regions, and time contexts interact in complex ways to determine the appropriate alarm threshold.
 
\subsection{Summary and Research Gap}
 
The literature shows that machine learning has made strong progress in predicting network faults, classifying alarms, and detecting anomalies. However, a clear gap remains when the works are examined against four criteria: support for multiple equipment vendors, direct threshold prediction, temporal adaptation to day of the week and time of day, and evaluation on real operational network data. Existing work on alarm management focuses on filtering and correlating alarms after they are raised, not on deciding what threshold should have raised them in the first place. Costa et al.~\cite{costa2009} and Jakobson \& Weissman~\cite{jakobson1993} discuss alarm classification and correlation, but they don't make any predictions or learn anything. On the other hand, Asres et al.~\cite{asres2022} learn from alarm history instead of thresholds and predict problematic ticket creation. Fault prediction work learns to predict failures based on performance indicators, but it fails to provide threshold values that can be used in an alarming system \cite{huang2019,murphy2019}. They predict whether a fault will occur but produce no threshold values. Deep learning work on time series has advanced the architectures available for sequential network data, but none of these models have been applied to threshold prediction as a target task \cite{vaswani2017,zhao2021,liu2024}. None of the reviewed works support multiple equipment vendors, adapt thresholds based on the time of day or day of the week, and evaluate on real operational network data all at the same time. No single prior work satisfies all four criteria simultaneously.

\section{Dataset, Preprocessing and Problem Formulation}
\label{sec:dataset-problem}
\subsection{Dataset Description}

\begin{figure}[h]
    \centering
    \includegraphics[width=0.8\textwidth]{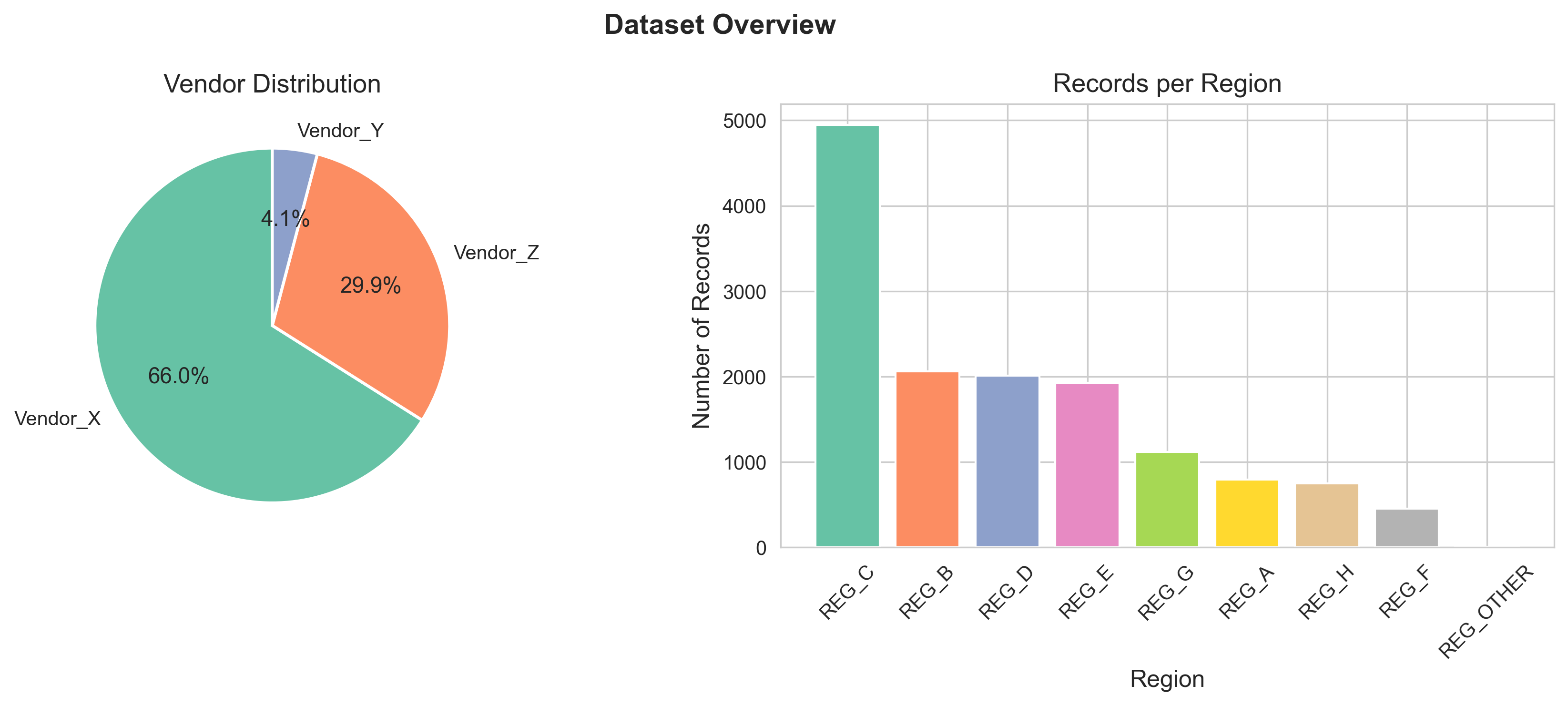}
    \caption{Vendor distribution and record count per geographic region.}
    \label{fig:Vendor distribution and record count per geographic region.}
\end{figure}

The dataset used in this study is an anonymized replica of real operational data from a functioning 4G mobile network in South Asia. The alarm management system creates a timestamped CSV file with an overview of all active cell alarms every ten minutes. The dataset covers ten sampled days spanning 14,054 cell day records from 10,648 unique cells across three equipment vendors and nine geographic regions. All cell identifiers and region codes have been replaced with generic labels and numeric values have been slightly perturbed using low magnitude Gaussian noise to prevent reverse identification while preserving the statistical distributions. Table~\ref{tab:dataset-characteristics} summarizes the key dataset properties and Figure~\ref{fig:Vendor distribution and record count per geographic region.} shows the vendor and regional composition.

\begin{table}[ht]
  \centering
  \caption{Dataset characteristics.}
  \label{tab:dataset-characteristics}
  \begin{tabular}{lc}
    \toprule
    Property & Value \\
    \midrule
    Total cell day records & 14,054 \\
    Unique cells & 10,648 \\
    Observation dates & 10 days \\
    Equipment vendors & 3 \\
    Geographic regions & 9 \\
    Alarm snapshots per day (approx.) & 144 \\
    Snapshot interval & 10 minutes \\
    Training samples (after expansion) & 201,984 \\
    Test samples & 135,120 \\
    \bottomrule
  \end{tabular}
\end{table}
 
\subsection{Preprocessing and Feature Engineering}

Each alarm file is loaded, the collection timestamp is extracted from the file details, and cell names are parsed using vendor specific logic. Each vendor uses a different convention for storing information. For Vendor X, the cell name is pulled directly from the alarmCustomAttr field. Vendor Y takes a different approach entirely, reconstructing the cell name from the source system name field based on a fixed character position pattern. Vendor Z tucks it inside the alarm attributes string, where it can be found under the cell name key. Files within the audit window are merged and duplicate alarms are removed by keeping the longest duration record per cell per alarm time. The data is then aggregated at the cell day level, computing total inactive time, fluctuation count, and hourly breakdowns for each of the 24 hours of the day. This produces 48 hourly features per record.

Additional features are derived from the date and time context including day of week, weekend flag, cyclical hour encodings, time block aggregations for night, morning, afternoon and evening periods, and one hot encoded vendor and region indicators. Log transforms are applied to inactive time and fluctuation count to reduce right skew. Prophet \cite{taylor2018} is used to extract a daily trend component and uncertainty interval from training dates, which are added as features to help the model anticipate high or low activity days. All continuous features are standardized using a scaler fitted on training data only. Each cell day record is then expanded across all 24 possible audit window start hours. This expansion means that for every combination of cell and date, the model treats the audit start times for each of the 24 hours as separate training samples. This design allows the model to learn how the optimal threshold changes depending on what time of day the operator initiates the audit. The expansion gives 201,984 training samples from seven days and 135,120 test samples from the three most recent days. The split is strictly time ordered so no future date appears in training, preventing any form of data leakage.

\subsection{Exploratory Data Analysis}
\label{subsec:eda}

Before building the models, we analyzed the dataset to understand the patterns that drive cell alarm behavior. This analysis directly shaped the feature engineering and label derivation decisions.

Table~\ref{tab:target-variable-summary} summarizes the key statistical properties of the two target variables. Inactive time is heavily right skewed with a median of 15.4 minutes but a 90th percentile of 236.8 minutes. Around 50 percent of cells show inactive time below 15 minutes per day while 19 percent exceed two hours, revealing a bimodal character that a single fixed threshold cannot serve well. Fluctuation counts are concentrated at very low values with a median of 2 and a maximum of 66. This extreme concentration at the lower end is important because it limits the ability of neural models to learn meaningful fluctuation thresholds, a finding discussed further in \Cref{sec:results-analysis}.
 
\begin{table}[ht]
  \centering
  \caption{Statistical summary of the two primary target variables.}
  \label{tab:target-variable-summary}
  \begin{tabular}{lcc}
    \toprule
    Statistic & Inactive Time (min) & Fluctuation Count \\
    \midrule
    Median & 15.4 & 2 \\
    75th percentile & 68.1 & 3 \\
    90th percentile & 236.8 & 4 \\
    Maximum & 1,256.1 & 66 \\
    Below floor ($\leq$15 min inactive / $\leq$1 fluctuation) & 49.6\% & 23.2\% \\
    Above 2 hours / above 10 & 19.1\% & N/A \\
    \bottomrule
  \end{tabular}
\end{table}
 
\begin{figure}[h]
    \centering
    \includegraphics[width=0.8\textwidth]{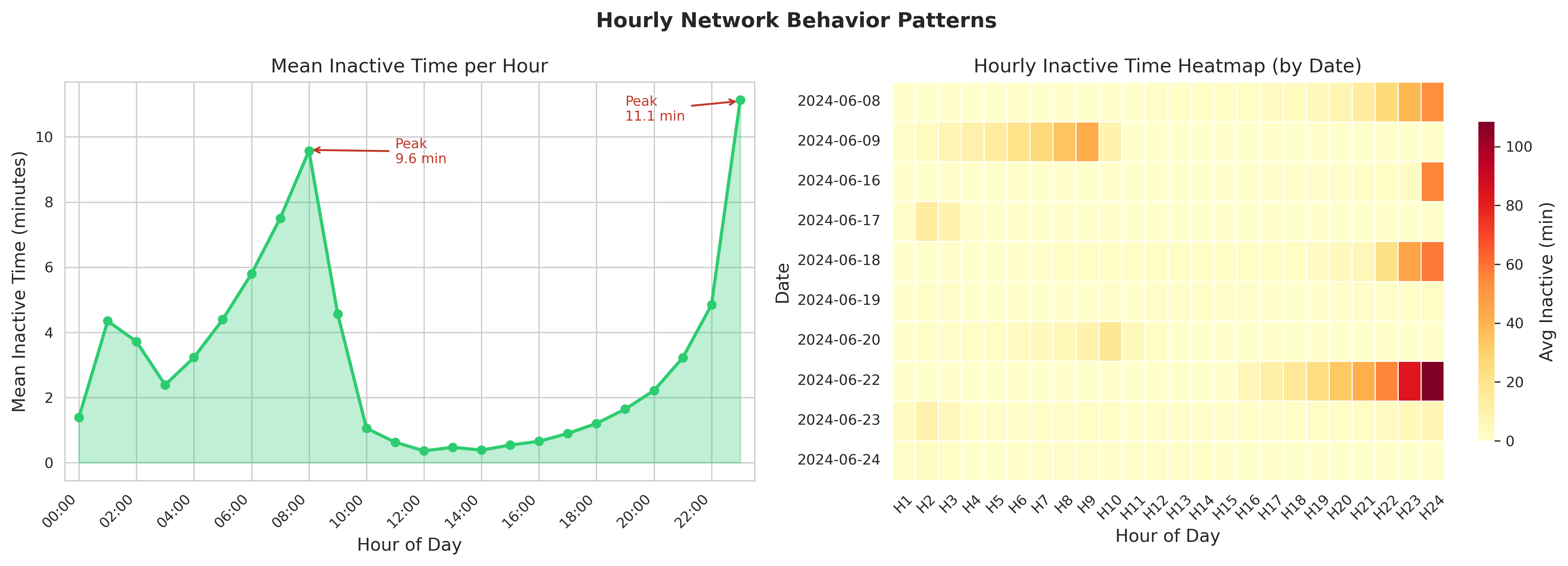}
    \caption{\textbf{Left}: mean inactive time per hour. \textbf{Right}: date wise heatmap of mean hourly inactive time.}
    \label{fig:Left: mean inactive time per hour. Right: date wise heatmap of mean hourly inactive time}
\end{figure}

As shown in Figure~\ref{fig:Left: mean inactive time per hour. Right: date wise heatmap of mean hourly inactive time}, the hourly breakdown reveals two distinct inactive time peaks. One is Hour 9 at 9.6 minutes mean and another is Hour 24 at 11.1 minutes mean. Hours 12 to 18 are nearly silent at under 0.4 minutes mean. Fluctuation count peaks at Hour 10. The date-wise heatmap panel in Figure~\ref{fig:Left: mean inactive time per hour. Right: date wise heatmap of mean hourly inactive time} shows that while this morning peak pattern is broadly consistent, individual dates deviate significantly in both timing and intensity. Weekend cells experience 4.3 times more inactive time than weekday cells on average. Vendor X has the highest mean inactive time at 85 minutes per day compared to 60 minutes for Vendor Z and 49 minutes for Vendor Y. Regional variation is also present, with REG D at 113 minutes and REG G at 36 minutes. These findings collectively confirm that start hour, day of week, vendor and region are all necessary predictors for adaptive threshold estimation.

\subsection{Problem Formulation}

We frame adaptive alarm threshold prediction as a supervised multi output regression problem. For a given cell, date and audit window start hour, the model predicts four threshold values. The input feature vector x contains 123 features grouped into six categories: temporal features encoding the date and start hour context (13 features), raw hourly inactive time and fluctuation values for all 24 hours (48 features), time block aggregations across night, morning, afternoon and evening periods (27 features), peak hour indicators and concentration scores (14 features), rolling and lag statistics from the previous day (4 features), and one hot encoded vendor and region indicators plus Prophet trend features (17 features).
\begin{center}
  \begin{tabular}{@{}lll@{}}
    $t_1$ & = \texttt{threshold\_hours} & (audit window length in hours, range 2 to 8) \\
    $t_2$ & = \texttt{threshold\_inactive\_min} & (maximum acceptable inactive time in minutes) \\
    $t_3$ & = \texttt{threshold\_fluctuation} & (maximum acceptable total fluctuation count) \\
    $t_4$ & = \texttt{each\_hour\_fluctuation} & (maximum acceptable fluctuation in any single hour)
  \end{tabular}
\end{center}

Since no ground truth labels exist in operational data, we derive them using a percentile based strategy. The window length t1 counts how many hours in the next eight hours show inactive time exceeding five minutes, clipped to the range two to eight. The inactive time threshold t2 is the 75th percentile of window inactive time multiplied by a sensitivity factor of 0.5 for evening hours (18:00 to 23:00), 1.0 for morning peak hours (07:00 to 11:00), and 1.25 for off peak hours, with a minimum floor of five minutes. Fluctuation thresholds t3 and t4 are the 90th percentile of window fluctuation counts with a floor of 1.

To check whether the label derivation introduces circular supervision, we held out 15 percent of cells entirely from the percentile computation and derived their labels using rules fitted on the remaining 85 percent only. A Kolmogorov-Smirnov test confirmed that the resulting label distributions were statistically indistinguishable from those produced using the full cell population across all four targets (threshold\_hours: p = 0.698; threshold\_inactive\_min: p = 0.718; both threshold\_fluctuation and each\_hour\_fluctuation: p = 1.000). This confirms both the percentile rules' stability and the absence of bias of the derived thresholds from including a cell in its own label computation, thereby justifying the use of the complete dataset in the final experiments.

\section{Proposed Models}
\label{sec:proposed-models}
\begin{figure}[h]
    \centering
    \includegraphics[width=0.8\textwidth]{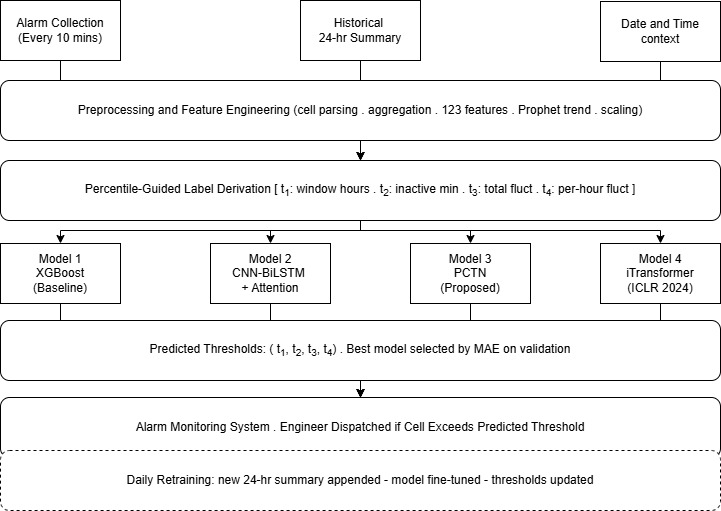}
    \caption{Adaptive Alarm Threshold Prediction Framework.}
    \label{fig:Adaptive Alarm Threshold Prediction Framework}
\end{figure}

Four models are evaluated in this study, covering a range from a classical tree based baseline to the novel PCTN architecture to a cutting-edge deep learning model. All models share the same 123 feature input and are trained to predict the four alarm thresholds simultaneously as a multi output regression task. The overall framework is shown in Figure~\ref{fig:Adaptive Alarm Threshold Prediction Framework}.

\subsection{Gradient Boosted Trees (XGBoost Baseline)}
Gradient Boosted Trees (GBT) \cite{friedman2001} serve as the classical baseline in this study. Here, XGBoost \cite{chen2016} creates a sequential array of decision trees, where every tree corrects the residual errors of all previous trees. It handles tabular data effectively without requiring feature scaling or architectural design, and provides feature importance rankings that aid interpretability.

Four separate XGBoost regressors are trained, one per target, using the pseudo Huber loss which is robust to the outlier inactive time values in the dataset:

\begin{equation}
  \ell_{\delta}(\hat{y}, y) = \delta^2 \left(\sqrt{1 + \left(\frac{y-\hat{y}}{\delta}\right)^2} - 1\right)
  \label{eq:pseudo-huber}
\end{equation}
 
A known limitation of tree based models is the inability to extrapolate beyond the training range of feature values. Because the test set includes a high-stress day and the feature values exceed those observed in training, test features are confined to the training range prior to prediction. This limitation and its effect on results are discussed in \Cref{sec:results-analysis}.

\subsection{CNN BiLSTM with Multi Head Attention}
\begin{figure}[h]
    \centering
    \includegraphics[width=0.8\textwidth]{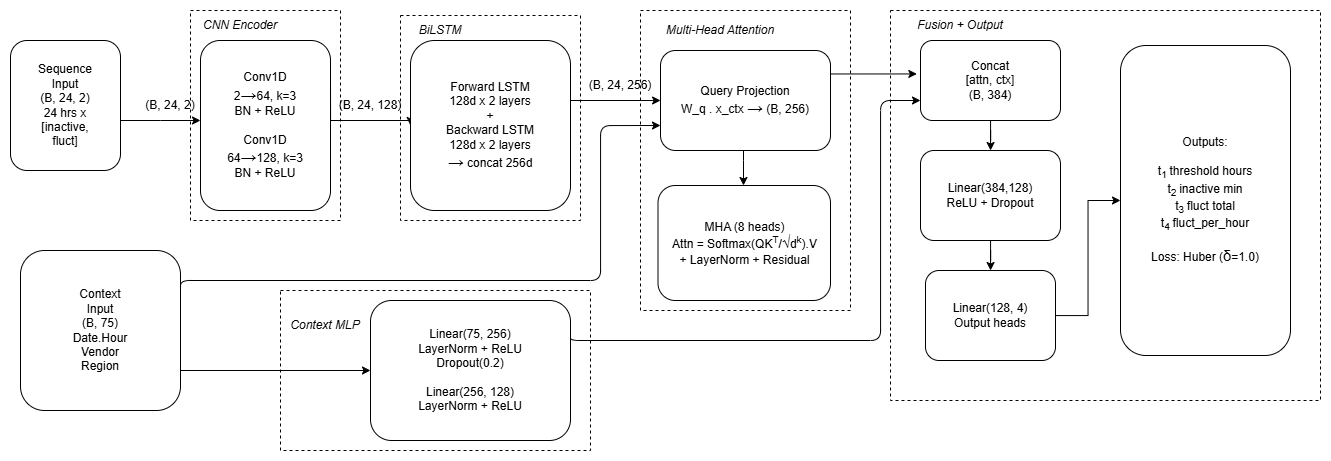}
    \caption{CNN-BiLSTM with Multi-Head Attention Architecture.}
    \label{fig:CNN-BiLSTM with Multi-Head Attention Architecture}
\end{figure}

The second model is a hybrid sequential architecture that treats the 24 hourly values as a temporal sequence and processes them through a convolutional encoder, a bidirectional LSTM, and context guided attention that highlights the most relevant hours for the given audit window. The full architecture is shown in Figure~\ref{fig:CNN-BiLSTM with Multi-Head Attention Architecture}.

The motivation for this design is that alarm behavior has both local structure where consecutive hours often share similar patterns and long-range dependencies. The morning peak at Hour 9 is partly predicted by the activity seen at Hour 2. CNN layers extract local hourly patterns \cite{zhao2021}, the BiLSTM captures dependencies in both temporal directions, and multi head attention allows the context vector, encoding the time of day, day of week, vendor, and region, to selectively weight the most relevant hours for the current audit window:
 
\begin{equation}
  A = \operatorname{Softmax}\!\left(\frac{QK^{\top}}{\sqrt{d_k}}\right)V
  \label{eq:attention-cnn-bilstm}
\end{equation}
 
where $Q$ is the context projected query, and $K$ and $V$ are projections of the BiLSTM output. The attended representation is combined with the context MLP output and sent to four output heads. The model is trained with Huber loss ($\delta=1.0$) using AdamW \cite{loshchilov2019} with cosine annealing and early stopping at patience 8.

\subsection{PCTN: Percentile Guided Contextual Threshold Network}
PCTN is the novel architecture proposed in this paper. Its central contribution is the separation of threshold prediction into two interpretable stages: learning the statistical distribution of network behavior for a given context, then deriving the threshold analytically from those learned parameters. The concept for the model is made clear by this design. An operator can inspect the predicted mean and variance to understand why a particular threshold was chosen.
 
\begin{figure}[h]
    \centering
    \includegraphics[width=0.8\textwidth]{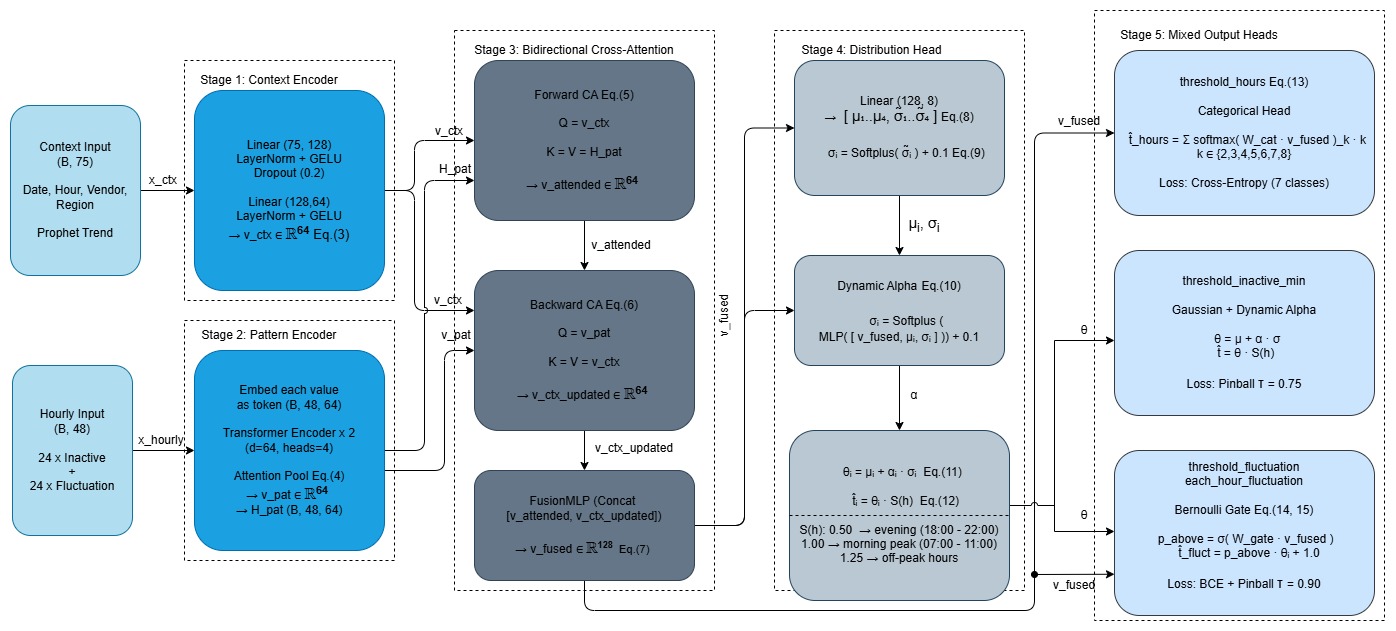}
    \caption{PCTN Architecture.}
    \label{fig:PCTN Architecture}
\end{figure}

Predicting a threshold directly as a regression target conflates two fundamentally different quantities: what the network typically does, and how far from that typical behavior should constitute an alert. By separating these into a distribution estimate and a scaling decision, PCTN makes its outputs both accurate and adjustable without retraining. The full architecture is shown in Figure~\ref{fig:PCTN Architecture}.

\subsubsection{Architecture}
 
The model takes two inputs. They are a context vector $x_{\textrm{ctx}}$ of 75 features (date, time, vendor, region, Prophet trend) and an hourly vector $x_{\textrm{hourly}}$ of 48 features (24 inactive + 24 fluctuation values). The model processes these through five stages.
 
Stage 1 encodes the context through a two layer MLP with LayerNorm and GELU activations \cite{hendrycks2016} to produce a context vector $v_{\textrm{ctx}} \in \mathbb{R}^{64}$:
 
\begin{equation}
  v_{\textrm{ctx}} = \textrm{GELU}\!\left(\textrm{LN}\!\left(W_2\,\textrm{GELU}\!\left(\textrm{LN}\!\left(W_1 x_{\textrm{ctx}}\right)\right)\right)\right)
  \label{eq:pctn-context-encoder}
\end{equation}
 
Stage 2 treats each of the 48 hourly values as an independent token and encodes them through a two layer Transformer encoder. Rather than using global average pooling, which discards positional information, an attention pooling layer learns to weight each hourly token by importance:
 
\begin{equation}
  v_{\textrm{pat}} = \sum \textrm{softmax}(w H_{\textrm{pat}})\,H_{\textrm{pat}}
  \label{eq:pctn-attn-pooling}
\end{equation}
 
where $w$ is a learnable vector. This ensures the model does not lose track of which specific hours tend to carry the most weight when making a threshold decision.
 
Stage 3 applies bidirectional cross attention. In the forward direction, the context queries the hourly pattern to figure out which hours matter most given the current time of day and day of week:
 
\begin{equation}
  v_{\textrm{attended}} = \textrm{CrossAttn}(Q=v_{\textrm{ctx}},\,K=V=H_{\textrm{pat}})
  \label{eq:pctn-crossattn-forward}
\end{equation}
 
When considering the backward approach, the hourly pattern gains understanding from the context, which helps the pattern evidence refine the context interpretation.
 
\begin{equation}
  v_{\textrm{ctx\_updated}} = \textrm{CrossAttn}(Q=v_{\textrm{pat}},\,K=V=v_{\textrm{ctx}})
  \label{eq:pctn-crossattn-backward}
\end{equation}
 
Both attended representations are concatenated and passed through a fusion layer:
 
\begin{equation}
  v_{\textrm{fused}} = \textrm{FusionMLP}(\textrm{Concat}[v_{\textrm{attended}},\,v_{\textrm{ctx\_updated}}])
  \label{eq:pctn-fusion}
\end{equation}
 
Stage 4 passes the fused representation through a distribution head that outputs the mean and variance for each of the four targets:
 
\begin{equation}
  [\mu_1,\ldots,\mu_4,\tilde{\sigma}_1,\ldots,\tilde{\sigma}_4] = W_{\textrm{dist}}\,v_{\textrm{fused}} + b
  \label{eq:pctn-dist-head}
\end{equation}
 
\begin{equation}
  \sigma_i = \textrm{Softplus}(\tilde{\sigma}_i) + 0.1
  \label{eq:pctn-sigma}
\end{equation}
 
The Softplus function ensures $\sigma_i > 0$ at all times. The mean $\mu_i$ captures what the network typically does in this context and $\sigma_i$ captures the variability around that expectation.
 
Stage 5 derives the final thresholds using target specific output heads that match the statistical structure of each target rather than applying a single Gaussian assumption to all four.
 
For threshold\_inactive\_min, which is a continuous target, a dynamic alpha parameter determines the conservatism of the threshold. Unlike a fixed scalar, alpha is a function of the predicted distribution parameters and the fused context:
 
\begin{equation}
  \alpha_i = \textrm{Softplus}(\textrm{MLP}([v_{\textrm{fused}},\mu_i,\sigma_i])) + 0.1
  \label{eq:pctn-alpha}
\end{equation}
 
\begin{equation}
  \theta_i = \mu_i + \alpha_i\sigma_i
  \label{eq:pctn-theta}
\end{equation}
 
\begin{equation}
  \hat{t}_i = \theta_i\,S(h)
  \label{eq:pctn-threshold}
\end{equation}
 
where $S(h)=0.5$ for evening hours (18:00 to 23:00), $1.0$ for morning peak (07:00 to 11:00), and $1.25$ for off peak hours. This sensitivity scaling enforces stricter thresholds during high customer impact periods.
 
For threshold\_hours, which takes discrete integer values from 2 to 8, a categorical head is used rather than Gaussian regression:
 
\begin{equation}
  \hat{t}_{\textrm{hours}} = \sum_{k=2}^{8} \textrm{softmax}(W_{\textrm{cat}}v_{\textrm{fused}})_k\,k
  \label{eq:pctn-hours}
\end{equation}
 
This calculates the expected value of the categorical distribution, which can be differentiated and optimized from start to finish using cross entropy loss. Applying Gaussian regression to a discrete ordinal target produces fractional outputs that do not correspond to any real operational value, which is why this head is necessary.
 
For threshold\_fluctuation and each\_hour\_fluctuation, where about 94 percent of values sit at the minimum floor of 1.0, a Bernoulli gate is applied before the Gaussian component:
 
\begin{equation}
  p_{\textrm{above}} = \sigma(W_{\textrm{gate}}v_{\textrm{fused}})
  \label{eq:pctn-gate}
\end{equation}
 
\begin{equation}
  \hat{t}_{\textrm{fluct}} = p_{\textrm{above}}\,\theta_i + 1.0
  \label{eq:pctn-fluct}
\end{equation}
 
The gate first predicts whether the cell is above the floor at all. If the gate output is near zero the prediction converges to the floor value. If above floor the Gaussian component predicts the magnitude. This two step structure allows the model to separately learn the boundary between normal and anomalous cells, rather than treating both as a single regression problem.
 
\subsubsection{Loss Function}
 
The four output heads use different loss functions matched to their statistical structure. For threshold\_hours the cross entropy loss over the seven integer classes is used. For threshold\_inactive\_min the pinball loss \cite{koenker1978} with $\tau=0.75$ is used:
 
\begin{equation}
  L_{\tau}(y,\hat{y}) = \tau\,\max(y-\hat{y},0) + (1-\tau)\,\max(\hat{y}-y,0)
  \label{eq:pinball-loss}
\end{equation}
 
For the two fluctuation targets a binary cross entropy loss is applied to the Bernoulli gate combined with a pinball loss at $\tau=0.90$ on the full output. The total loss is the mean over all five components.
 
The asymmetry in the pinball loss is operationally important. Taking $\tau=0.75$ as an example, the model gets penalized three times as heavily for underestimating the threshold as it does for overestimating it. This matches the real world cost asymmetry: a threshold set too low misses real faults and denies customers service, while a threshold set too high generates unnecessary engineer callouts, which is costly but less harmful.

\subsection{iTransformer}
\begin{figure}[h]
    \centering
    \includegraphics[width=0.8\textwidth]{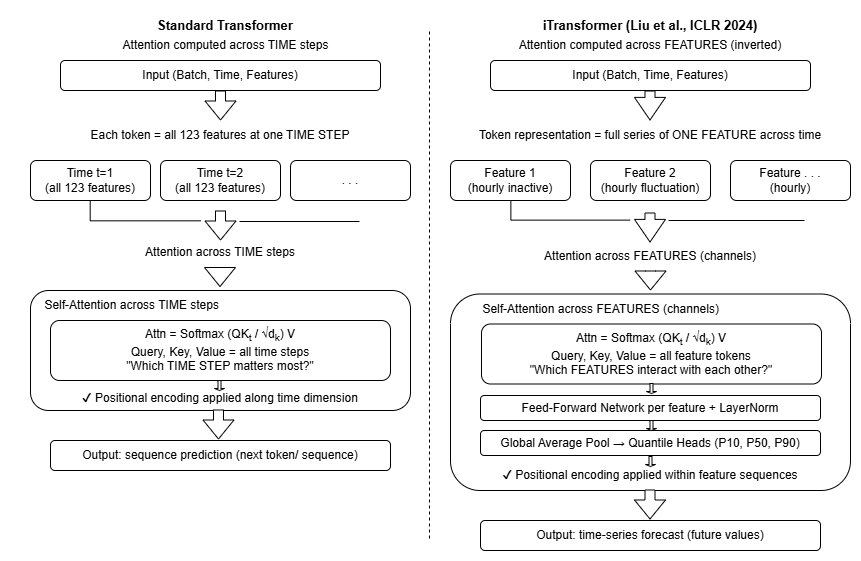}
    \caption{Standard Transformer vs iTransformer: The Feature-Inversion Concept.}
    \label{fig:Standard Transformer vs iTransformer: The Feature-Inversion Concept}
\end{figure}

  The fourth model adapts the iTransformer~\cite{liu2024}, a 2024 ICLR Spotlight paper, to the multi output threshold prediction task. The iTransformer challenges the conventional Transformer design for multivariate data and introduces a simple but powerful architectural inversion. Instead of forming tokens from all features at each time step, it forms tokens from all time steps of each feature. The comparison between standard and inverted Transformers is shown in Figure~\ref{fig:Standard Transformer vs iTransformer: The Feature-Inversion Concept}.

In alarm monitoring, the question is not which time step matters most. All steps are already aggregated into hourly features. The question is which features interact most strongly to determine risk. So, applying attention across features rather than across time is operationally justified.

Each feature is embedded into a 128 dimensional vector ($d_{\mathrm{model}}=128$). The inverted self attention follows the same scaled dot-product formulation as Eq.~\ref{eq:attention-cnn-bilstm}, but $Q$, $K$, and $V$ are now projections of the feature token representations rather than time step representations. This is applied for $L=4$ encoder layers. We adapt the iTransformer for multi output threshold prediction by replacing the original forecasting head with three quantile output heads ($P10$, $P50$, $P90$). The $P90$ output is used as the final threshold prediction, making the model conservative by design. The combined quantile loss is:
 
\begin{equation}
  L = \frac{1}{3}\left[ L_{0.10}(y,q_{10}) + L_{0.50}(y,q_{50}) + L_{0.90}(y,q_{90}) \right]
  \label{eq:itransformer-quantile-loss}
\end{equation}
 
The three quantile outputs also provide a calibrated uncertainty range at inference time. The gap between $q_{10}$ and $q_{90}$ signals how confidently the model can predict the threshold for a given context.

\subsection{Training Configuration}

Each model is trained using the same seven-day training set and tested over the same three-day period. Table~\ref{tab:training-config} summarizes the configuration for each model.

\begin{table}[ht]
  \centering
  \caption{Training configuration summary.}
  \label{tab:training-config}
  \begin{tabular}{lcccll}
    \toprule
    Model & Parameters & Optimizer & Epochs & Loss Function & Hardware \\
    \midrule
    XGBoost & N/A & Gradient & 300 & Pseudo-Huber & CPU \\
    CNN-BiLSTM & 1,153,732 & AdamW & 50 & Huber ($\delta=1.0$) & GTX 1650 Ti \\
    PCTN & 142,261 & AdamW & 60 & CE + Pinball + BCE & GTX 1650 Ti \\
    iTransformer & 818,508 & AdamW & 60 & Quantile (3-head) & T4 GPU \\
    \bottomrule
  \end{tabular}

  \vspace{0.3em}
  \small\textit{Note:} CE refers to Cross-Entropy and BCE refers to Binary Cross-Entropy.
\end{table}

An important design consideration for all four models is their compatibility with daily incremental retraining. For XGBoost, the model is fully retrained on the accumulated dataset each day, which is feasible because tree training is fast even on large tabular datasets. For the three neural models, only a fine tuning step of five to ten epochs is required when new daily data arrives, rather than full retraining. This is possible because model weights learned from previous days provide a strong starting point that needs only minor modifications to capture new trends. This design means that the system is always getting better, and as more operational data is collected, threshold predictions get more accurate and better at managing the network's unique behaviors.

\section{Experimental Setup}
\label{sec:experimental-setup}
All models are trained on data from the seven earliest dates and evaluated on the three most recent dates, preserving temporal order to prevent data leakage. Models 2 and 3 are trained on an NVIDIA GTX 1650 Ti GPU, Model 4 on an NVIDIA T4 GPU via Google Colab, and Model 1 on CPU. Each experiment uses a fixed random seed of 42 for reproducibility. No hyperparameter search is performed and the configurations in Table~\ref{tab:training-config} are used directly. No validation set tuning was performed beyond the early stopping criterion applied during training. All reported metrics are only calculated using the held-out test set, and the temporal ordering of the split ensures that no future Information influenced any model during training or hyperparameter selection.
 
Performance is measured using Mean Absolute Error (MAE), Root Mean Squared Error (RMSE), and the coefficient of determination $R^2$. Metrics are reported per target and averaged across all four targets for overall model comparison. For each test sample, we compute the average absolute error across all four targets to produce a single per-sample error scalar, and the Wilcoxon test \cite{wilcoxon1945} is then applied to compare the distributions of these per-sample errors between two models. For the iTransformer, the P90 quantile output is used as the threshold prediction, reflecting a conservative policy where 90 percent of normally behaving cells fall below the predicted threshold.

\begin{figure}[h]
    \centering
    \includegraphics[width=0.8\textwidth]{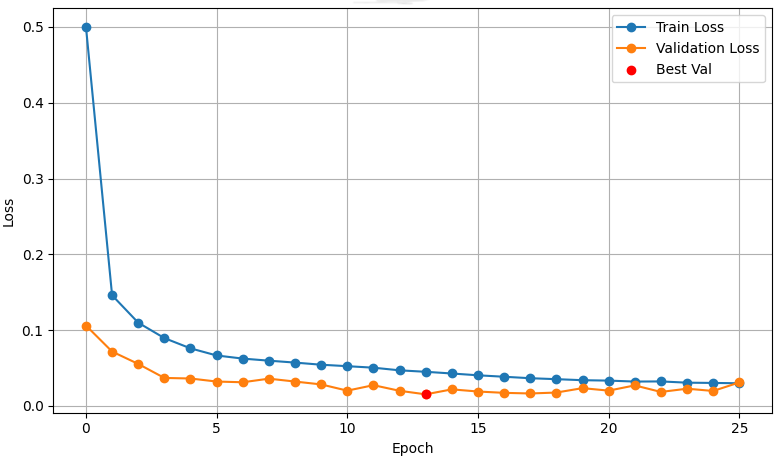}
    \caption{PCTN Training and Validation Loss.}
    \label{fig:PCTN Training and Validation Loss}
\end{figure}

Figure~\ref{fig:PCTN Training and Validation Loss} shows the training and validation loss curves for PCTN. The best validation loss epoch is highlighted in red. This shows that the model converged correctly and that the early stopping criterion worked to avoid overfitting.

\section{Results and Analysis}
\label{sec:results-analysis}
\subsection{Overall Performance Comparison}
 
Table~\ref{tab:overall-performance} reports MAE, RMSE, and $R^2$ for all four models across all four targets. The proposed PCTN achieves the best overall average $R^2$ of 0.733 and the best average MAE of 0.290, outperforming iTransformer ($R^2=0.436$), CNN-BiLSTM ($R^2=0.324$), and XGBoost ($R^2=-31.03$) with only 142,261 parameters, 83\% fewer than iTransformer. XGBoost’s failure on threshold\_hours and threshold\_inactive\_min is discussed in \Cref{subsec:xgboost-fails}. PCTN achieves the best result on three of the four individual targets. According to the Wilcoxon signed-rank test, every pairwise comparison between PCTN and the three baseline models is statistically significant at $p<0.001$.

\begin{table}[ht]
  \centering
  \caption{Overall model performance across all targets. Best values per row are marked with $\ast$.}
  \label{tab:overall-performance}
  \begin{tabular}{llcccc}
    \toprule
    Metric & Target & XGBoost & CNN-BiLSTM & PCTN & iTransformer \\
    \midrule
    MAE & threshold\_hours & 1.2010 & 0.2361 & 0.0224$\ast$ & 0.0843 \\
     & threshold\_inactive\_min & 143.3483 & 1.3773 & 1.0638$\ast$ & 1.1328 \\
     & threshold\_fluctuation & 0.0137$\ast$ & 0.0838 & 0.0348 & 0.0532 \\
     & each\_hour\_fluctuation & 0.0137$\ast$ & 0.0734 & 0.0380 & 0.0578 \\
     & Average & 36.1442 & 0.4427 & 0.2898$\ast$ & 0.3320 \\
    \midrule
    RMSE & threshold\_hours & 2.5481 & 0.4439 & 0.1470$\ast$ & 0.1542 \\
     & threshold\_inactive\_min & 143.9753 & 7.4073 & 5.6015 & 5.2594$\ast$ \\
     & threshold\_fluctuation & 0.0853$\ast$ & 0.2598 & 0.1702 & 0.2643 \\
     & each\_hour\_fluctuation & 0.0853$\ast$ & 0.2611 & 0.1695 & 0.2645 \\
     & Average & 36.6735 & 2.0930 & 1.5221 & 1.4856$\ast$ \\
    \midrule
    $R^2$ & threshold\_hours & -12.4598 & 0.5915 & 0.9552$\ast$ & 0.9507 \\
     & threshold\_inactive\_min & -113.4414 & 0.6971 & 0.8268 & 0.8473$\ast$ \\
     & threshold\_fluctuation & 0.8930$\ast$ & 0.0080 & 0.5739 & -0.0271 \\
     & each\_hour\_fluctuation & 0.8930$\ast$ & -0.0019 & 0.5776 & -0.0282 \\
     & Average & -31.0288 & 0.3237 & 0.7334$\ast$ & 0.4357 \\
    \midrule
    Parameters & -- & N/A & 1,153,732 & 142,261 & 818,508 \\
    \bottomrule
  \end{tabular}
\end{table}

To contextualise these results against current operational practice, a naive fixed-threshold baseline was simulated by setting each threshold to its training-set mean value for all test samples. This baseline approximates manually configured static thresholds in operational networks. It produced an average MAE of 0.499 and an average $R^2$ of 0.0 by construction. These results confirm that even the weakest neural model (CNN-BiLSTM, average MAE $=0.443$) improves over static fixed thresholds, while the proposed PCTN (average MAE $=0.290$) reduces prediction error by 42\% relative to this baseline.

\subsection{Performance on the Operationally Critical Targets}

The two targets that matter most for real alarm system design are threshold\_hours and threshold\_inactive\_min because they directly control when and how long a cell is monitored before an alarm fires. PCTN achieves $R^2=0.9552$ on threshold\_hours and $R^2=0.8268$ on threshold\_inactive\_min. The strong result on threshold\_hours is driven by the categorical head, which matches the discrete nature of this target, combined with attention pooling that preserves which specific hours are active, which is the primary signal determining window length. On threshold\_inactive\_min, iTransformer achieves $R^2=0.8473$, marginally ahead of PCTN by 2 percentage points, but requires 818,508 parameters. PCTN delivers comparable accuracy at 83\% lower cost. For deployments where model size, inference latency, or hardware budget is constrained, PCTN is the practical choice without meaningful accuracy loss.

\subsection{Fluctuation Targets and the Bernoulli Gate}
\label{subsec:fluctuation-gate}

PCTN achieves $R^2=0.5739$ and $0.5776$ on the two fluctuation targets, while all other neural models produce near-zero or slightly negative values. XGBoost achieves $R^2=0.893$ on these targets but fails on the other two as discussed in \Cref{subsec:xgboost-fails}. The PCTN result is entirely due to the Bernoulli gate. As established in \Cref{subsec:eda}, 94.3\% of fluctuation threshold values in the test set sit at the minimum floor of 1.0. Standard neural regression learns to predict values near 1.0 everywhere, giving low MAE but near-zero $R^2$ because it never identifies genuinely anomalous cells. The Bernoulli gate decomposes prediction into two decisions: whether the cell is above the floor, then what magnitude to predict if it is. This explicit modeling of the binary structure is what separates PCTN from the three baselines on this problem.

The near-zero or slightly negative $R^2$ values for iTransformer on fluctuation targets confirm a practically important principle: model capacity alone does not solve structurally imbalanced targets. An architecture with 5.8 times more parameters than PCTN cannot handle floor concentration without an explicit structural solution. When deploying alarm monitoring systems in networks where cell fluctuation behavior is highly imbalanced, output-head design matters more than model size.

\subsection{Why XGBoost Fails on Two Targets}
\label{subsec:xgboost-fails}

XGBoost produces $R^2=-12.46$ and $-113.44$ on threshold\_hours and threshold\_inactive\_min, despite being the strongest model for fluctuation targets. The cause is feature extrapolation failure. The test set contains the highest-stress day in the dataset, where the feature \texttt{Window\_Inactive\_Sum} reached values of 15.3, more than double the training maximum of 7.5. Tree-based models cannot extrapolate beyond the range observed during training. When a test sample falls outside all leaf boundaries, the model returns the nearest leaf prediction, producing severely inflated threshold estimates.

This finding has a direct practical implication for production deployment: tree-based models for this task are reliable under stable conditions but brittle under novel stress events without precedent in training data. Neural models generalize more gracefully because their continuous weight representations interpolate between learned patterns. In environments where occasional extreme network events are expected, a neural model provides a safer and more robust choice even when accuracy on typical days is comparable to a tree-based alternative.

\subsection{PCTN Interpretability: Dynamic Alpha}
After training, the dynamic alpha values across the test set are summarized in Table~\ref{tab:dynamic-alpha}.

\begin{table}[ht]
  \centering
  \caption{Dynamic alpha statistics across the test set.}
  \label{tab:dynamic-alpha}
  \begin{tabular}{lcc}
    \toprule
    Target & Mean $\alpha$ & Std $\alpha$ \\
    \midrule
    threshold\_hours & 4.073 & 0.768 \\
    threshold\_inactive\_min & 14.183 & 2.678 \\
    threshold\_fluctuation & 0.688 & 0.121 \\
    each\_hour\_fluctuation & 0.144 & 0.077 \\
    \bottomrule
  \end{tabular}
\end{table}

\begin{figure}[h]
    \centering
    \includegraphics[width=0.8\textwidth]{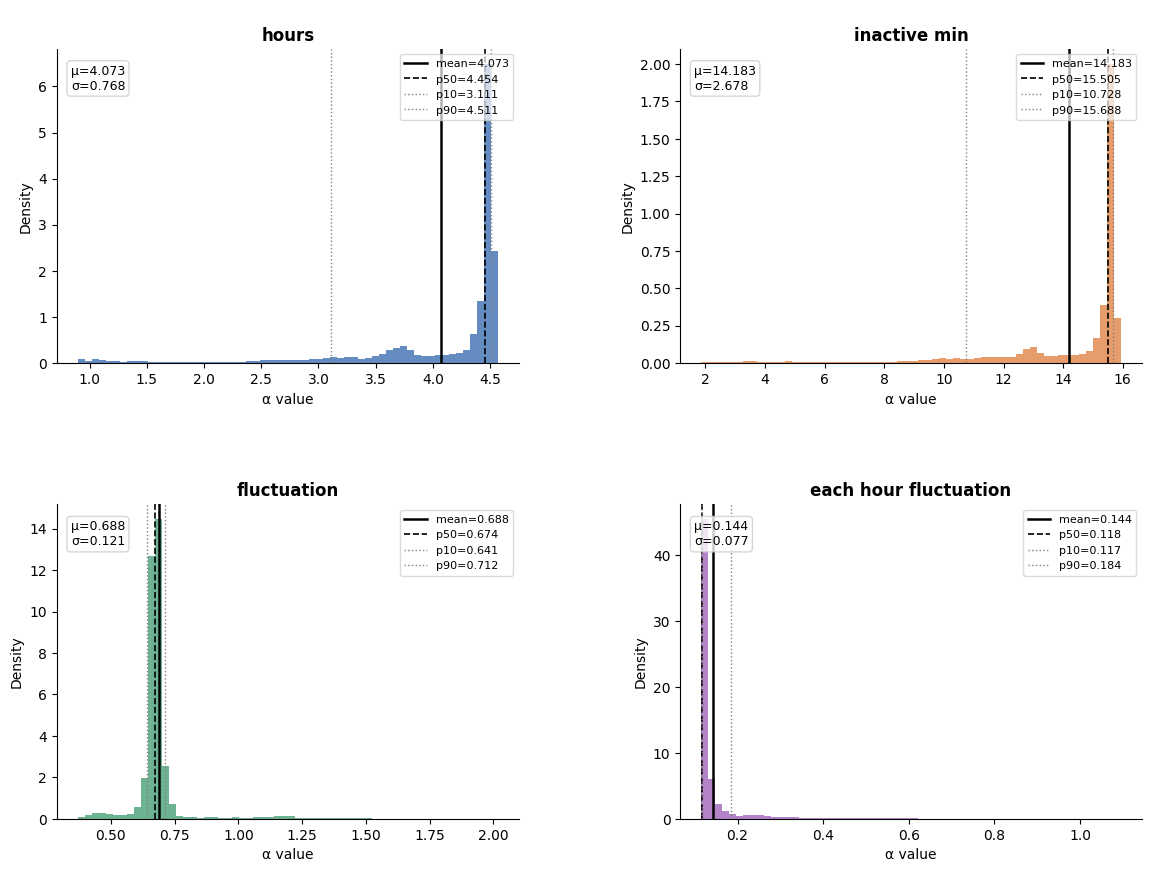}
    \caption{Dynamic Alpha ($\alpha$) Distribution Across Test Samples.}
    \label{fig:dynamic-alpha-distribution}
\end{figure}

Figure~\ref{fig:dynamic-alpha-distribution} visualizes the full per-sample $\alpha$ distributions across all 135,120 test instances. The non-trivial standard deviations for threshold\_hours and threshold\_inactive\_min confirm that alpha varies meaningfully across samples rather than collapsing to a constant, validating the dynamic-alpha design. For threshold\_inactive\_min, alpha ranges from approximately 8.8 to 22.2, indicating substantially more conservative behavior on high-variability days than on quiet days. For the fluctuation targets, the narrower alpha distributions (std $=0.121$ and $0.077$) reflect deliberate division of labor with the Bernoulli gate introduced in \Cref{subsec:fluctuation-gate}. Since the gate models whether a fluctuation value exceeds the floor, it absorbs most sample-level variation for these targets, leaving alpha to apply a consistent upward shift on above-floor samples.

No other model in this comparison gives an operator a direct, adjustable handle on conservatism for each cell and context. An operator managing a high-risk maintenance window can increase effective alpha for specific cells without retraining, a capability unique to PCTN's distribution-based design.

\subsection{iTransformer Uncertainty Quantification}

The iTransformer produces three quantile outputs at inference time: $P10$, $P50$, and $P90$. Table~\ref{tab:quantile-spread} shows the average $P10$-to-$P90$ spread per target on the test set.

\begin{table}[ht]
  \centering
  \caption{Average quantile spread ($P10$ to $P90$) for iTransformer predictions.}
  \label{tab:quantile-spread}
  \begin{tabular}{lc}
    \toprule
    Target & Average Spread \\
    \midrule
    threshold\_hours & 0.19 hours \\
    threshold\_inactive\_min & 1.02 minutes \\
    threshold\_fluctuation & 0.01 \\
    each\_hour\_fluctuation & 0.02 \\
    \bottomrule
  \end{tabular}
\end{table}

The narrow spread on fluctuation targets reflects floor dominance discussed in Section~6.4 rather than genuine model confidence. The wider spread on threshold\_inactive\_min (1.02 minutes) reflects real variability in how inactive-time thresholds should respond to different network contexts. In deployment, audit windows with a spread significantly above the average can be flagged for secondary review before threshold application, providing a built-in safety layer at no additional computational cost since all three quantile values are produced in a single forward pass.

\section{Limitations and Future Work}
\label{sec:limitations-future}
\subsection{Limitations}

The most immediate limitation of this study is the size of the dataset. Ten days of operational data from a single national network, while sufficient to demonstrate the feasibility of the approach and validate the model architectures, is a narrow window of network behavior. It cannot fully capture seasonal variation, the effect of major network events such as hardware upgrades or regional outages, or the long term changes in traffic patterns that operators see over months and years. A model trained on ten days may perform well within the observed behavioral range but will encounter patterns it has never seen before as more time passes. So the results should be interpreted as a proof of concept rather than a production ready system.

The dataset covers only 4G network infrastructure. Modern deployments are more and more integrating 5G and 4G cells at the same sites, while 5G cells have unique radio behaviors, traffic behavior, and fault behaviors. Incorporating 5G data into the framework will involve extra feature engineering to identify 5G specific indicators to apply the system to the operations of next generation networks.

Fluctuation thresholds remain partially unresolved. Although the Bernoulli gate in PCTN significantly improved performance over all other neural models, $R^2 = 0.57$ still leaves a meaningful prediction error. Networks where fluctuation events are more frequent and diverse are expected to yield better results, but this cannot be confirmed without additional data from different network types.

Finally, the label derivation strategy is based on percentile rules applied to the same data the model trains on. In an ideal setting, ground truth labels would come from expert annotation of historical thresholds that produced good operational outcomes, which would provide stronger validation of the derived supervision signals.

\subsection{Future Work}

The most practical near term improvement is extending the dataset. Adding three to six months of operational data across different seasons would expose the model to a wider range of network behavior and reduce the risk of distributional failure on high stress days. Collecting data from multiple national networks, even if anonymized, would allow evaluation of how well the models generalize across different operators and infrastructure profiles.

For the fluctuation threshold problem, a hybrid deployment strategy is recommended for immediate production use. The inactive time and window duration thresholds should be handled by the trained PCTN model, while the fluctuation thresholds are computed using a simple daily percentile rule updated from the last seven days of data. This removes the need to solve the floor concentration problem entirely and is deployable with the existing architecture.

The most significant long term improvement opportunity is adapting PCTN to a few shot learning regime using meta learning. In the current design, PCTN requires a full dataset from the target network to train. This means a new operator must collect weeks of data before the model can be used. Training PCTN under a Model Agnostic Meta Learning objective \cite{finn2017} would allow it to adapt to a new network from as few as five to ten days of data, directly removing the data scarcity barrier that currently limits real world adoption.

A complementary direction that requires no architectural change is inference-time calibration of the dynamic $\alpha$ values. The temperature parameter $\alpha$ controls the stochasticity of the policy \cite{haarnoja2018}, while adaptive thresholding methods adjust decision boundaries based on recent observations without gradient-based optimization. If a cell has triggered alarms at a higher rate than the network average in the past week, its effective $\alpha$ can be automatically adjusted to enforce a more conservative threshold for that cell. This per-cell personalization acts as a form of few-shot adaptation that exploits the interpretability of PCTN’s design in a way that would be difficult to achieve with black-box models.

\section{Conclusion}
\label{sec:conclusion}
Mobile network operators currently set alarm thresholds manually and keep them fixed, irrespective of the time of day, day of the week, or the actual behavior of the network. This means the same threshold that is appropriate at midnight is also applied during morning peak hours when millions of customers are active, leading to missed faults during busy periods and unnecessary engineer callouts at quiet times. This paper addressed that problem by framing adaptive alarm threshold prediction as a supervised machine learning task for the first time.

We introduced a percentile guided label derivation strategy that creates training supervision from raw network behavior data where no ground truth threshold labels exist, and we proposed PCTN, the Percentile Guided Contextual Threshold Network, which learns the statistical distribution of network behavior for a given context and derives thresholds analytically from those learned parameters. PCTN is the first model designed specifically for this task. The model uses different output layers to match each type of prediction. It uses a categorical layer for discrete durations, a Gaussian layer for continuous inactive time, and a Bernoulli layer for fluctuation patterns. This structural alignment with the data is what allows PCTN to outperform architecturally larger models.

Experiments on an anonymized dataset of 10,648 cells across three equipment vendors and nine geographic regions from a real 4G network show that PCTN consistently outperforms all three baselines across the operationally critical targets, achieving the best overall average performance while using 83 percent fewer parameters than the best performing deep learning alternative. The model also learns to change its thresholds based on the condition of the network instead of a pre-determined set of rules, and operators can review and manually change these adjustments for particular cells without any model retraining.

The framework is built for continuous daily retraining, as the thresholds adapt as the network behavior changes over time, instead of being locked at a value determined months ago. As network complexity increases and 5G deployments expand, the ability to automatically adapt alarm thresholds from operational data to maintain quality of service without a proportional increase in engineering effort will become increasingly important.

\section{Credit Authorship Contribution Statement}

Ayon Roy: Conceptualization, Methodology, Software, Formal analysis, Investigation, Writing — original draft, Writing — review and editing. \\
Sadman Sharif: Visualization, Writing — review and editing. \\
Shiva Prasad Sarker: Data curation, Software.

\section{Declaration}
Throughout the work, authors used AI-assisted tools to fix grammatical error in writing, debug code, and organize the manuscript. The authors generated and validated the remainder of the scientific content, experimental designs, model architectures, analysis, and conclusions. The full responsibility concerning the authenticity of the published work is accepted by the authors.
The authors state that there are no other financial interests or affiliations that are relevant to the research or are perceived to potentially bias the work described in this manuscript.
\section{Acknowledgements}

Data supporting this study's findings can be obtained from the corresponding author on reasonable request. The authors thank the network operations team for providing access to the operational alarm data that made this study possible.
\pagebreak
\bibliographystyle{plain}
\bibliography{references}

\end{document}